\documentclass{article}

\usepackage{arxiv}

\usepackage[utf8]{inputenc} % allow utf-8 input
\usepackage[T1]{fontenc}    % use 8-bit T1 fonts
\usepackage{hyperref}       % hyperlinks
\usepackage{url}            % simple URL typesetting
\usepackage{booktabs}       % professional-quality tables
\usepackage{amsfonts}       % blackboard math symbols
\usepackage{nicefrac}       % compact symbols for 1/2, etc.
\usepackage{microtype}      % microtypography
\usepackage{lipsum}

\title{The probability flow in the Stock market and Spontaneous symmetry breaking in Quantum Finance}

\author{
 Ivan Arraut \\
  Lee Shau Kee School of Business and Administration\\
 The Open University of Hong Kong,\\
  30 Good Shepherd Street, Homantin, Kowloon \\
  \texttt{ivanarraut05@gmail.com} \\
 \And
  Jo$\tilde{\bf a}$o Alexandre Lobo Marques \\
  FBL, University of Saint Joseph\\
  Estrada Marginal da Ilha Verde, 14-17, Macao, China\\
  \And
   Sergio Gomes\\
  FBL, University of Saint Joseph\\
  Estrada Marginal da Ilha Verde, 14-17, Macao, China\\ \\
  \texttt{sergio.gomes@usj.edu.mo} \\
}

\begin{document}
\maketitle

\begin{abstract}
The Spontaneous Symmetry breaking in Quantum Finance considers the martingale condition in the stock market as a vacuum state if we express the financial equations in the Hamiltonian form. The original analysis for this phenomena ignores completely the kinetic terms in the neighborhood of the minimal of the potential terms. This is correct in most of the cases. However, when we deal with the Martingale condition, it comes out that the kinetic terms can also behave as potential terms and then reproduce a shift on the effective location of the vacuum (Martingale). In this paper we analyze the effective symmetry breaking patterns and the connected vacuum degeneracy for these special circumstances. Within the same scenario, we analyze the connection between the flow of information and the multiplicity of martingale states, providing in this way powerful tools for analyzing the dynamic of the stock market. 
\end{abstract}

% keywords can be removed
\keywords{Martingale condition \and Vacuum condition \and spontaneous symmetry breaking \and degenerate vacuum}

\section{Introduction}

When we express the Financial equations of the Stock market in the Hamiltonian form, the flow of information through the market can be quantified by analyzing the flow of probability. The financial Hamiltonians are in general non-Hermitian and then they do not preserve the information. However, under some special combination of the free-parameters of the market, the financial Hamiltonian can become Hermitian, preserving then the information (probability). The flow of information in the market, is connected with the definition of the martingale or equilibrium state, which is the equilibrium state where there is no flow of information. In probability theory, a martingale process is the one where the future expectation value of a random variable is just the present value \cite{1}. In Quantum Finance, the martingale condition corresponds to a risk-neutral evolution, consistent with the most basic concepts taken from probability theory \cite{2}. A neutral evolution, is free from any possibility of arbitrage \cite{2, 3}. Arbitrage in finance gives the possibility of investors to operate in different markets \cite{3}. For example, in general it is possible for a broker to buy sharings in New York and then sell them in Hong Kong, getting in this way, some income from the differences in the prices in both markets over the same product. This can be done in a market outside the equilibrium condition (outside martingale). In addition, only big corporations can get benefits with some income from arbitrage due to the high volume in their inversions. Individuals cannot receive enough earnings by using arbitrage due to the fee charges in transactions which would cancel any possibility of income \cite{3}. The arbitrage process helps the market to arrive to the equilibrium state. Once the market is in this equilibrium condition, any possibility or arbitrage is lost. It is at this point when we have a martingale process with a risk-neutral evolution. In fact, the existence of a martingale condition is known as {\bf the fundamental theorem of finance} \cite{3, 4}. In a previous paper, some of the authors formulated the Spontaneous Symmetry breaking in Quantum Finance \cite{OurEPL}. Within this formalism, the Martingale condition (state) appears as a vacuum condition which becomes degenerate under some circumstances \cite{3, OurEPL}. In the most general sense, this vacuum condition is non-unique for the Black-Scholes (BS) and the Merton-Garman (MG) cases when we explore the symmetries under change of prices and the symmetry under changes in volatility for the MG case. These symmetries come out to be spontaneously broken \cite{7} (their generators do not annihilate the vacuum state), except for some combination of parameters, which guarantee the vacuum (martingale state) to be unique ( non-degenerate). For the regime analyzed in \cite{OurEPL}, the same mentioned combination of parameters guarantee the no-flow of information through the boundaries of the system. This is an interesting connection between spontaneous symmetry breaking and flow of information, already perceived in a different context in \cite{Winchi}. In \cite{OurEPL} it was also formulated an extended martingale condition for the case of the MG equation, which considers the martingale state not only as a function of the prices of the stock, but also on the stochastic volatility. For this case again the vacuum, taken as the martingale state, is degenerate and then the corresponding symmetries are spontaneously broken. For this extended concept of Martingale, under the regimes explored in \cite{OurEPL}, the conditions for the vacuum to be single come out to be not only the necessity of no-flow of information through the boundaries of the system, but also the absence of any white noise coming from the stochastic volatility. These two conditions guarantee the martingale state (vacuum) to be single, restoring then the symmetries for the vacuum state. 
Another interesting situation was analyzed in \cite{OurEPL}. It corresponds to an ideal case where for the MG and the BS case, additional non-derivative terms are included such that the martingale condition is still satisfied. These potential terms are different to those analyzed before by some other authors \cite{3, 5}. The potential terms added in \cite{OurEPL} are also different to the standard case studied in \cite{6}, where a double slit constraint was explored. Indeed, the potential terms analyzed in \cite{OurEPL} correspond to collective decisions or collective behavior. Although the results obtained in \cite{OurEPL} are correct, they were always focused on the analysis of situations ignoring the kinetic contributions in the neighborhood of the Martingale state. This corresponds to the strong-field regime to be defined in this paper. This regime, although valid, does not represent the whole scenario. For this reason, in this paper we also explore other regimes by extending the results obtained in \cite{OurEPL}. We then explore situations where the kinetic terms behave as additional potential terms (weak-field regime and intermediate regimes). This certainly happens in the reality when we consider the Martingale condition in the BS equation as well as for the MG equation. Indeed, the kinetic terms cannot be ignored in general around the neighborhood of the martingale state when we consider its standard definition. Still we can say that there are regimes where the results obtained in \cite{OurEPL} are valid and regimes where the additional results obtained in this paper represent a more accurate picture of the reality. Finally, in this paper we fully connect the notions of spontaneous symmetry breaking and flow of information (probability) in the stock market. The paper is organized as follows: In Sec. (\ref{Sec1}), we describe the BS equation and we express it in its Hamiltonian form. In Sec. (\ref{MGequationsec}), we explain the MG equation and its corresponding Hamiltonian. In Sec, (\ref{Sec2}), we explain the meaning of a martingale condition from the classical perspective in Finance. In Sec. (\ref{Sec3}), we explain why the martingale condition, for the evolution of an Option, is equivalent to a vacuum condition from the perspective of Quantum Mechanics. In Sec. (\ref{Sec4}), we introduce some potential terms in the financial Hamiltonians and we analyze under which conditions they preserve the martingale condition. In Sec. (\ref{NS1}), we extend the results obtained in \cite{OurEPL} where it was shown that the symmetries under a changes in the prices are spontaneously broken. This situation occurs, except for some particular combinations of the parameters of the Hamiltonian. The same situation appears for the symmetries under changes of volatility for the MG equation if we extend the notion of martingale states in order to include the volatility as one of the variables. In the cases studied in this paper, we consider the kinetic term contributions which were ignored in \cite{OurEPL}. This means that in this paper we explore both, weak and strong regimes for the Quantum field representing the series expansion of the martingale state. The martingale state then corresponds to the vacuum condition for this quantum field denoted by $\phi_{vac}$. Its explicit result depends on the regime under analysis as well as the order of the series expansion when we express the Hamiltonian as a function of Quantum fields. In Sec. (\ref{Secnew}), we analyze the details about the extended martingale condition which depends not only on the prices of the Options but also on the stochastic volatility. This is the section where the explicit results for the MG case are analyzed. In Sec.... Probability flow

Finally, in Sec. (\ref{Sec5}), we conclude.

\section{The Black-Scholes equation}   \label{Sec1}

The stock price $S(t)$ is normally taken as a random stochastic variable evolving in agreement to a stochastic differential equation given by

\begin{equation}   \label{rainbowl}
\frac{dS(t)}{dt}=\phi S(t)+\sigma SR(t).
\end{equation}   
Here $\phi$ is the expected return of the security, $R(t)$ is the Gaussian white noise with zero mean and $\sigma$ is the volatility \cite{3}. Note that this simple equation contains one derivative term on the left-hand side and non-derivative terms on the right-hand side. The fundamental analysis of Black and Scholes, exclude the volatility such that we can guarantee the evolution of the price of the stock with certainty \cite{9}. In this way, by imposing $\sigma=0$, we obtain a simple solution for the equation (\ref{rainbowl}) as

\begin{equation}
S(t)=e^{\phi t}S(0).    
\end{equation}
The possibility of arbitrage is excluded if we can make a perfect hedged portfolio. In this sense, any possibility of uncertainty is excluded and we can analyze the evolution of the price free of any white noise \cite{3}. We can consider the following portfolio

\begin{equation}   \label{forever}
\Pi=\psi-\frac{\partial \psi}{\partial S}S.
\end{equation}
This is a portfolio where an investor holds the option and then {\it short sells} the amount $\frac{\partial \psi}{\partial S}$ for the security $S$. By using the Ito calculus (stochastic calculus) \cite{3}, it is possible to demonstrate that

\begin{equation}   \label{BS}
\frac{d\Pi}{dt}=\frac{\partial \psi}{\partial t}+\frac{1}{2}\sigma^2S^2\frac{\partial^2 \psi}{\partial S^2}.
\end{equation}
Here the change in the value of $\Pi$ does not have any uncertainty associated to it \cite{3}. The random term has disappeared due to the choice of portfolio. Since here we have a risk-free rate of return for this case (no arbitrage) \cite{Europe, Epjons}, then the following equation is satisfied

\begin{equation}   \label{hedged}
\frac{d\Pi}{dt}=r\Pi.
\end{equation} 
If we use the results (\ref{forever}) and (\ref{BS}), together with the previous equation, then we get

\begin{equation}   \label{BSeq}
\frac{\partial \psi}{\partial t}+rS\frac{\partial \psi}{\partial S}+\frac{1}{2}\sigma^2S^2\frac{\partial^2\psi}{\partial S^2}=r\psi.
\end{equation}
This is the Black-Scholes equation \cite{Op2, Op3, Merton2}, which is independent of the expectations of the investors, defined by the parameter $\phi$, which appears in eq. (\ref{rainbowl}). In other words, in the Black-Scholes equation, the security (derivative) price is based on a risk-free process. The Basic Assumptions of the Black-Scholes equation are:\\
1). The spot interest rate $r$ is constant.\\
2). In order to create the hedged portfolio $\Pi$, the stock is infinitely divisible, and in addition it is possible to short sell the stock.\\
3). The portfolio satisfies the no-arbitrage condition.\\
4). The portfolio $\Pi$ can be re-balanced continuously.\\
5). There is no fee for transaction.\\ 
6). The stock price has a continuous evolution. \\

\subsection{Black-Scholes Hamiltonian formulation}

We will explain how the eq. (\ref{BSeq}), can be expressed as an eigenvalue problem after a change of variable. The resulting equation will be the Schr\"odinger equation with a non-Hermitian Hamiltonian. For eq. (\ref{BSeq}), consider the change of variable $S=e^x$, where $-\infty<x<\infty$. In this way, the BS equation becomes

\begin{equation}   \label{BSHamiltonian2}
\frac{\partial\psi}{\partial t}=\hat{H}_{BS}\psi,
\end{equation} 
where we have defined the operator  

\begin{equation}   \label{BSHamiltonian}
\hat{H}_{BS}=-\frac{\sigma^2}{2}\frac{\partial^2}{\partial x^2}+\left(\frac{1}{2}\sigma^2-r\right)\frac{\partial}{\partial x}+r.
\end{equation}
as the BS Hamiltonian. Note that the resulting Hamiltonian is non-Hermitian since $\hat{H}\neq\hat{H}^+$ \cite{5}. In addition, note that since the spot interest rate $r$ is constant, then the potential term is just a constant term. This means that the vacuum condition is trivial for this case. Under the BS Hamiltonian, the evolution in time of the Option is non-unitary in general (in addition, the Hamiltonian non-necessarily obeys the $PT$ symmetry). This means that the probability is not necessarily preserved in time, although it is certainly well-defined and its total value is equal to one. In general, there are some cases in ordinary Quantum Mechanics, as well as in Quantum Field Theory, where it is interesting to explore non-Hermitian Hamiltonians (Lagrangians) \cite{8}. Based on the previous explanations, we cannot expect the financial market to obey unitarity. The reason for this is simply because the market is not a closed system and there are many external factors influencing its behavior as for example it is the amount of people and organizations trading at some instant of time. Then the assumption of unitarity makes no sense at all and the Hamiltonian must be non-Hermitian. In this paper however, when we add some potential terms to the BS and MG equations, we will impose the Hermiticity condition on them. When the symmetry is spontaneously broken and we are working at the vacuum level, all the terms in the original BS equation become irrelevant, giving then importance to (only) the potential terms. In this way we can follow the standard formalism suggested in \cite{8}. The same conclusions apply to any other equation only containing kinetic terms as it is the case of the Merton-Garman (MG) equation to be analyzed shortly. Inside the MG case however, the conclusions about what is happening during the breaking of the symmetry of the market (spontaneously) are more relevant than in the BS case. We will return back to this point later in this paper.        

\section{The Merton-Garman equation: Preliminaries and derivation}   \label{MGequationsec}

We can consider a more general case where the security and the volatility are both stochastic. In such a case, the market is incomplete \cite{3}. Although several stochastic processes have been considered for modeling the case with stochastic volatility \cite{45}, here we consider the generic case, defined by the set of equations \cite{3}

\begin{eqnarray}   \label{corona5}
\frac{dS}{dt}=\phi Sdt+S\sqrt{V}R_1\nonumber\\
\frac{dV}{dt}=\lambda+\mu V+\zeta V^\alpha R_2.
\end{eqnarray}
Here the volatility is defined through the variable $V=\sigma^2$ and $\phi$, $\lambda$, $\mu$ and $\zeta$ are constants \cite{51}. The Gaussian noises $R_1$ and $R_2$, corresponding to each of the variables under analysis, are correlated in the following form

\begin{equation}   \label{corona3}
<R_1(t')R_1(t)>=<R_2(t')R_2(t)>=\delta(t-t')=\frac{1}{\rho}<R_1(t)R_2(t')>.    
\end{equation}
Here $-1\leq\rho\leq1$, and the bra-kets $<AB>$ correspond to the correlation between $A$ and $B$. If we consider a function $f$, depending on the Stock price, the time, as well as on the white noises; with the help of the Ito calculus, it is possible to derive the total derivative in time of this function as

\begin{eqnarray}
\frac{df}{dt}=\frac{\partial f}{\partial t}+\phi S\frac{\partial f}{\partial S}+(\lambda+\mu V)\frac{\partial f}{\partial V}+\frac{\sigma^2S^2}{2}\frac{\partial^2f}{\partial S^2}+\rho V^{1/2+\alpha}\zeta\frac{\partial^2f}{\partial S\partial V}+\frac{\zeta^2V^{2\alpha}}{2}\frac{\partial^2f}{\partial V^2}+\sigma S\frac{\partial f}{\partial S}R_1+\zeta V^\alpha\frac{\partial f}{\partial V}R_2.    
\end{eqnarray}
This equation can be expressed in a more compact form which separates the stochastic terms from the non-stochastic ones as follows

\begin{equation}   \label{HK}
\frac{df}{dt}=\Theta+\Xi R_1+\psi R_2.    
\end{equation}
 Here we have defined 
 
\begin{eqnarray}
\Xi=\sigma S\frac{\partial f}{\partial S}, \;\;\;\;\; \;\;\;\;\; \;\;\;\;\;\psi=      
\zeta V^\alpha\frac{\partial f}{\partial V},\nonumber\\
\Theta=\frac{\partial f}{\partial t}+\phi S\frac{\partial f}{\partial S}+(\lambda+\mu V)\frac{\partial f}{\partial V}+\frac{\sigma^2S^2}{2}\frac{\partial^2f}{\partial S^2}+\rho V^{1/2+\alpha}\zeta\frac{\partial^2f}{\partial S\partial V}+\frac{\zeta^2V^{2\alpha}}{2}\frac{\partial^2f}{\partial V^2},
\end{eqnarray} 
 keeping in this way the notation used in \cite{3} for convenience. 
 
 \subsection{Derivation of the Merton-Garman equation}
 
 If we consider two different options defined as $C_1$ and $C_2$ on the same underlying security with strike prices and maturities given by $K_1$, $K_2$, $T_1$ and $T_2$ respectively. It is possible to create a portfolio 
 
 \begin{equation}
\Pi=C_1+\Gamma_1C_2+\Gamma_2S.     
 \end{equation}
 If we consider the result (\ref{HK}), then we can define the total derivative with respect to time as for the folio as
 
 \begin{equation}
\frac{d\Pi}{dt}=\Theta_1+\Gamma_1\Theta_2+\Gamma_2\phi S+\left(\Xi_1+\Gamma_1\Xi_2+\Gamma_2\sigma S\right)R_1+\left(\psi_1+\Gamma_1\psi_2\right)R_2.     
 \end{equation}
Note that this result is obtained after recognizing $f(t)=C_1$ or $f(t)=C_2$ in eq. (\ref{HK}) when it corresponds. It has been demonstrated that even in this case of stochastic volatility, it is still possible to create a hedged folio and then at the end we arrive again to the condition (\ref{hedged}), after finding special constraints for $\Gamma_1$ and $\Gamma_2$ such that the white noises are removed. The solution for $\Pi$ is a non-trivial one for this case, and then it requires the definition of the parameter

\begin{eqnarray}   \label{beta}
\beta(S, V, t, r)=\frac{1}{\partial C_1/\partial V}\left(\frac{\partial C_1}{\partial t}+(\lambda+\mu V)\frac{\partial C_1}{\partial S}+\frac{VS^2}{2}\frac{\partial^2 C_1}{\partial S^2}+\rho V^{1/2+\alpha}\zeta\frac{\partial^2 C_1}{\partial S\partial V}+\frac{\zeta^2V^{2\alpha}}{2}\frac{\partial^2 C_1}{\partial V^2}-rC_1\right)\nonumber\\
=\frac{1}{\partial C_2/\partial V}\left(\frac{\partial C_2}{\partial t}+(\lambda+\mu V)\frac{\partial C_2}{\partial S}+\frac{VS^2}{2}\frac{\partial^2 C_2}{\partial S^2}+\rho V^{1/2+\alpha}\zeta\frac{\partial^2 C_2}{\partial S\partial V}+\frac{\zeta^2V^{2\alpha}}{2}\frac{\partial^2 C_2}{\partial V^2}-rC_2\right)
\end{eqnarray}
This parameter does not appear for the case of the BS equation. Indeed $\beta$ in the MG equation is defined as the market price volatility risk because the higher its value is, the lower is the intention of the investors to risk. Take into account that in the MG equation the volatility is a stochastic variable. Since the volatility is not traded in the market, then it is not possible to make a direct hedging process over this quantity \cite{3}. In this way, when we have stochastic volatility, it is necessary to consider the expectations of the investors. This effect appears through the parameter $\beta$. It has been demonstrated in \cite{66} that the value of $\beta$ in agreement with eq. (\ref{beta}) is a non-vanishing result. In general, it is always assumed that the risk of the market (in price) has been included inside the MG equation. The MG equation is then obtained by rewriting the equation (\ref{beta}) in the form

\begin{equation}   \label{MGE}
\frac{\partial C}{\partial t}+rS\frac{\partial C}{\partial S}+(\lambda+\mu V)\frac{\partial C}{\partial V}+\frac{1}{2}VS^2\frac{\partial^2 C}{\partial S^2}+\rho\zeta V^{1/2+\alpha}S\frac{\partial^2 C}{\partial S\partial V}+\zeta^2 V^{2\alpha}\frac{\partial^2 C}{\partial V^2}=rC,   
\end{equation}
where the effects of $\beta$ now appear contained inside the modified parameter $\lambda$ in this equation. In other words, we have shifted the parameter $\lambda\to \lambda-\beta$ in eq. (\ref{MGE}). Later in this paper, we will express this equation in the Hamiltonian form, which is the ideal one for understanding the concept of spontaneous symmetry breaking in Quantum Finance. 

\subsection{Hamiltonian form of the Merton-Garman equation}

The previously analyzed MG equation can be formulated as a Hamiltonian (eigenvalue) equation. We can define a change of variable defined as

\begin{eqnarray}   \label{range}
S=e^x,\;\;\;\;\;-\infty<x<\infty,\nonumber\\
\sigma^2=V=e^y,\;\;\;\;\;-\infty<y<\infty,
\end{eqnarray}
and then the MG equation (\ref{MGE}) becomes \cite{3, 5B, 70}

\begin{equation}
\frac{\partial C}{\partial t}+\left(r-\frac{e^y}{2}\right)\frac{\partial C}{\partial x}+\left(\lambda e^{-y}+\mu-\frac{\zeta^2}{2}e^{2y(\alpha-1)}\right)\frac{\partial C}{\partial y}+\frac{e^y}{2}\frac{\partial^2 C}{\partial x^2}+\rho\zeta e^{y(\alpha-1/2)}\frac{\partial^2 C}{\partial x\partial y}+\zeta^2 e^{2y(\alpha-1)}\frac{\partial^2 C}{\partial y^2}=rC.  
\end{equation}
If we express this equation as an eigenvalue problem in the same form as in eq. (\ref{BSHamiltonian2}) for the BS case, then we have

\begin{equation}
\frac{\partial C}{\partial t}=\hat{H}_{MG}C,   
\end{equation}
with the MG Hamiltonian defined as

\begin{equation}  \label{MGHamilton}
\hat{H}_{MG}=-\frac{e^y}{2}\frac{\partial^2 }{\partial x^2}-\left(r-\frac{e^y}{2}\right)\frac{\partial }{\partial x}-\left(\lambda e^{-y}+\mu-\frac{\zeta^2}{2}e^{2y(\alpha-1)}\right)\frac{\partial}{\partial y}-\rho\zeta e^{y(\alpha-1/2)}\frac{\partial^2 }{\partial x\partial y}-\zeta^2 e^{2y(\alpha-1)}\frac{\partial^2 }{\partial y^2}+r.   
\end{equation}
Exact solutions for the MG equation have been found for the case $\alpha=1$ in \cite{5B} by using path-integral techniques. The same equation has been solved in \cite{45} for the case $\alpha=1/2$ by using standard techniques of differential equations. Note that the equation has two degrees of freedom. Later we will see that when we have spontaneous symmetry breaking, it becomes irrelevant to know the exact solution of this equation.   

\section{The martingale condition in finance}   \label{Sec2}

The martingale condition is required for having a risk-neutral evolution for the price of an Option. This means that the price of a financial instrument is free of any possibility of arbitrage. In probability theory, the risk-free evolution is modeled inside a stochastic process. Assume for example $N+1$ random variables $X_i$, with a joint probability distribution defined as $p(x_1, x_2, ..., x_{N+1})$. Then the martingale process is simply defined as the condition under which

\begin{equation}   \label{martingale}
E[X_{n+1}\vert x_1, x_2, ..., x_n]=x_n,    
\end{equation}
is satisfied \cite{3}. Note that $E[X_i]$ is the expectation value of the random variable. Eq. (\ref{martingale}) suggests that the expected value of a subsequent observation of a random variable is simply the present value. For the purpose of this paper, the random variables correspond to the future prices of the stock given by $S_1$, $S_2$, ..., $S_{N+1}$, which are defined at different times $t_1$, $t_2$, ..., $t_{N+1}$. We can then apply the same martingale condition to the stocks if we make the corresponding discounts in order to compare prices defined at different moments \cite{2, 3}. We can assume that the future value of an equity is defined as $S(t)$. If there is a free-risk evolution of the discounted price defined as

\begin{equation}   \label{securdiscount}
e^{-\int_0^tr(t')dt'}S(t).  
\end{equation}
Then the value follows the martingale process \cite{42}. In this way the conditional probability for the present price is the actual value given by $S(0)$. The martingale condition can then be expressed as \cite{3}

\begin{equation}   \label{interpret}
S(0)=E\left[e^{-\int_0^tr(t')dt'}S(t)\vert S(0)\right],   
\end{equation}
and this result is general. Equivalent expressions have been used for the analysis of the evolution of forward rates \cite{3}. The importance of martingales is analyzed in \cite{80}. The interpretation of eq. (\ref{interpret}) is clear. The left-hand side is just the present price of the security. The right-hand side is the expected value of the discounted price of the security at the time $t$. Discounted means that the quantity evaluated at the time $t$ has to be extrapolated to the present value. Both quantities must be equivalent under the martingale condition.     

\section{The martingale condition as a vacuum condition for a Hamiltonian}   \label{Sec3}

The previous section dealt with the martingale condition. This section deals with the equivalent formulation of the same principle but from the perspective of The Hamiltonian formulation. Consider as before an option on a security $S=e^x$ that matures at time $T$ with the corresponding pay-off function $g(x)$. In this way we can describe the risk-free evolution of the option as 

\begin{equation}   \label{thisone}
C(t, x)= \int_{-\infty}^\infty dx'<x\vert e^{-(T-t)\hat{H}}\vert x'>g(x').    
\end{equation}
By using the previous definition of martingales, for this case we have

\begin{equation}
S(t)=E\left[e^{-(t_*-t)r}S(t_*)\vert S(t)\right].    
\end{equation}
If we introduce $S(x)$ (the price of the security) in eq. (\ref{thisone}), then under the martingale condition, we have 

\begin{equation}
S(t, x)= \int_{-\infty}^\infty dx'<x\vert e^{-(t_*-t)\hat{H}}\vert x'>S(x').    \end{equation}
This equation can be re-expressed in Dirac notation as 

\begin{equation}
<x'\vert S>= \int_{-\infty}^\infty dx'<x\vert e^{-(t_*-t)\hat{H}}\vert x'><x'\vert S>.    
\end{equation}
If we take the base $\vert x'>$ as a complete set of states, then the condition $\hat{I}=\int dx'\vert x'>< x'\vert$ ($\hat{I}$ is the identity matrix) is satisfied and then the previous expression is simplified as

\begin{equation}
\vert S>=e^{-(t_*-t)\hat{H}}\vert S>.    
\end{equation}
Then there is no Hamiltonian (time) evolution for the state $\vert S>$ under the previous conditions. It also comes out that the Hamiltonian annihilates the same state as follows

\begin{equation}   \label{MartHam}
\hat{H}\vert S>=0.    
\end{equation}
Interestingly, the BS Hamiltonian given in eq. (\ref{BSHamiltonian}) as well as the MG Hamiltonian giving in eq. (\ref{MGHamilton}) satisfy the martingale condition in the form defined in eq. (\ref{MartHam}). 

\section{Non-derivative terms introduced in the financial Hamiltonians}   \label{Sec4}

It is possible to introduce potential terms to the BS equation as well as to the MG one as it is explained in \cite{3}. It has been demonstrated that the martingale condition can still be maintained if the potential satisfies some special conditions. In general, a potential term will appear as

\begin{equation}   \label{non-hermBS}
\hat{H}_{BS, MG}^{eff}=\hat{H}_{BS, MG}+\hat{V}(x),    
\end{equation}
with the potential term $\hat{V}$ containing non-derivative terms depending on the security $S$. Since we usually have a change of variables in the Hamiltonian formulation, this functional dependence is indirect. In the previous equation, $\hat{H}_{BH, MG}^{eff}$ is the effective Hamiltonian including the potential contribution. Some barrier options as well as some path-dependent options admit the inclusion of potential terms for their deep understanding \cite{Potential}. On the other hand, for the case of the Black-Scholes Hamiltonian, the martingale condition is maintained if the potential appears in the Hamiltonian in the following form \cite{3}

\begin{equation}   \label{effblack}
\hat{H}_{BS}^{eff}=-\frac{\sigma^2}{2}\frac{\partial^2}{\partial x^2}+\left(\frac{1}{2}\sigma^2-V(x)\right)\frac{\partial}{\partial x}+V(x).    
\end{equation}
Then an effective Hamiltonian expressed in this way can be used for pricing the option. The discount in these general cases depends on the price of the option itself. Then the security discount defined in eq. (\ref{securdiscount}) is modified as \cite{3}

\begin{equation}
e^{-\int_0^tr(t')dt'}S(t)\to e^{-\int_0^tV(x(t'))dt'}S(t).    
\end{equation}
In \cite{3} it is argued that the usual discounting of a security using the spot interest rate $r$ is determined by the argument of no arbitrage involving fixed deposits in the money market account. Studies about viable potentials matching with the reality of the market are under analysis. It is important to notice that the Hamiltonian (\ref{effblack}) can be converted to a Hermitian operator by using a similarity transformation as has been reported in \cite{3}. Then we can define

\begin{equation}   \label{wecandefine}
\hat{H}_{BS}^{eff}=e^s\hat{H}_{Herm}e^{-s}.    
\end{equation}
Here the Hermitian Hamiltonian is defined as

\begin{equation}   \label{Alex}
\hat{H}_{Herm}=-\frac{\sigma^2}{2}\frac{\partial^2}{\partial x^2}+\frac{1}{2}V'(x)+\frac{1}{2\sigma^2}\left(V+\frac{1}{2}\sigma^2\right)^2,    
\end{equation}
and $s=x/2-(1/\sigma^2)\int^x_0 dy V(y)$. This result can be obtained by replacing (\ref{effblack}) and (\ref{Alex}) in eq. (\ref{wecandefine}). From the Hermitian Hamiltonians, it is possible to construct a complete basis and then we can find real eigenvalues associated to this Hamiltonian. Note in particular that in the Black-scholes case $V(x)=r$ is constant. It is a simple task to demonstrate that the Hermitian Hamiltonian obtained by similarity transformation can be also expressed as

\begin{equation}
\hat{H}_{Herm}=e^{\alpha x}\left(-\frac{\sigma^2}{2}\frac{\partial^2}{\partial x^2}+\gamma\right)e^{-\alpha x},  
\end{equation}
with 

\begin{equation}
\gamma=\frac{1}{2\sigma^2}\left(r+\frac{1}{2}\sigma^2\right)^2\;\;\;\;\;\alpha=\frac{1}{\sigma^2}\left(\frac{1}{2}\sigma^2-r\right).    
\end{equation}
Among the trivial examples of potentials already analyzed in the literature, we find the Down-and-Out barrier option, where the stock price has to be over some minimal value, below which it becomes worthless. This behavior can be guarantee with an infinite potential barrier boundary condition imposed for the value of the corresponding price. This case can be worked out directly from the non-Hermitian Hamiltonian defined in eq. (\ref{non-hermBS}). Another example of potential corresponds to the Double-Knock-Out barrier option, where it is easier to work with the Hermitian Hamiltonian part as it is defined in eq. (\ref{wecandefine}) but including the potential part as follows

\begin{equation}   \label{standarddefinitio}
\hat{H}_{DB}=\hat{H}_{BS}+\hat{V}(x)=e^s\left(\hat{H}_{Herm}+\hat{V}(x)\right)e^{-s}.    
\end{equation}
This definition is used for analyzing cases where the stock has to be maintained between a maximal and a minimal value. Note that the definition (\ref{standarddefinitio}) can be also used for the analysis of the Down-and-Out barrier if we focus on the non-Hermitian part (the term in the middle of the equation) with the corresponding potential. Note that these examples of potentials representing real situations are trivial cases. More details about these examples can be found in \cite{3}.   

\section{Deeper analysis for the Black-Scholes and the Merton-Garman equation}   \label{NS1}

If we analyze the martingale condition, we can notice that its interpretation as a vacuum condition is not perfect. The reason is that although the Hamiltonian annihilates the martingale state $\vert S>$, as can be seen from eq. (\ref{MartHam}), the momentum operator corresponding to the prices of the options ($\hat{p}_x$) does not annihilate the vacuum perfectly. This can be seen from their definitions as follows 

\begin{equation}
\hat{p}_x\vert S>=e^x\vert S>.    
\end{equation}
This means that the symmetry under translations of the prices, carried out from the security $S$ is spontaneously broken. An exception is the case where $S\to0$ and then $x\to-\infty$ as can be seen from eq. (\ref{range}). This however, would give us a trivial value for the security as $S=0$. For general values, the symmetry under translations of prices is spontaneously broken. This means that the different values of $S$ represent different possible vacuums or ground states. Different vacuums however, would have different amount of information because under the ideal conditions, all the information of the market is stored inside the prices. 
The action of $\hat{p}_x$ over $\vert S>$, then maps one vacuum toward another one defined through the selected value $x$. Such operation could be seen as a rotation in a complex plane if we make the transformation of variables $x\to i n\theta$. In such a case, we have 

\begin{equation}
\hat{p}_x\vert S>=e^{in\theta}\vert S>=\vert S'>\neq\vert S>. 
\end{equation}
Here $n$ is just a number and $\theta$ is a dimensionless phase. After this change of variable, the action of $\hat{p}_x$ is to map one vacuum into another one through a rotation defined by the phase $\theta$. This condition shows the vacuum degeneracy. The same condition has different meanings depending on whether we consider the BS or the MG equation. This important detail about the Martingale condition deserves more attention. 

\subsection{Reinterpretation of the Martingale condition}                                                           
We can re-define the martingale state $S(x, t)$ as a Quantum Field by doing the following change of variable $S(x, t)=e^x=\sum_{n=0}^\infty x^n/n!=\sum_{n=0}^\infty\phi^n(x, t)$, as it was proposed in \cite{OurEPL}. Then we can express the BS equation as a function of the Quantum Field $\phi(x, t)$ by replacing this definition inside eq. (\ref{BSHamiltonian}), obtaining in this way

\begin{equation}  \label{allpotentials}
\hat{H}_{BS}S(x, t)=-\frac{\sigma^2}{2}n(n-1)\phi^{n-2}+\left(\frac{1}{2}\sigma^2-r\right)n\phi^{n-1}+r\phi^n(x, t)=V(\phi)=0,    
\end{equation}
which is valid in the neighborhood of the martingale state (vacuum). Here we omit the sum symbol because the comparison has to be done term by term in the series. This means that we have one equation for each term in the series expansion for $\phi$. In eq. (\ref{allpotentials}), we can see that all the terms in the Hamiltonian can now be considered to be potential terms. Note that the result (\ref{allpotentials}) is valid if $d\phi/dx=\sum_n\phi/n$. In some regimes (especially inside the weak-field approximation), the kinetic terms in eq. (\ref{allpotentials}) play a fundamental role in the identification of the vacuum state. The vanishing condition in eq. (\ref{allpotentials}) is a natural consequence of the Martingale state definition, which being a vacuum condition, is annihilated by the Hamiltonian. This is the case because the symmetries under time-translation are not spontaneously broken. We can solve the martingale condition in eq. (\ref{allpotentials}), obtaining then 

\begin{equation}   \label{newmar}
\phi_{vac}=\frac{n}{2}\left(1-\frac{\sigma^2}{2r}\right)\left(1\pm\sqrt{1+\frac{8r\sigma^2}{(\sigma^2-2r)^2}\frac{(n-1)}{n}}\right), \;\;\;n\neq0.
\end{equation}
This is the definition of the martingale state in the sense of the Quantum Field $\phi$, as a function of the free-parameters of the theory. Note that for $n=0$, the result (\ref{newmar}) is not well-defined, and then we impose the condition $n\neq0$. Another special case is $n=1$, for which, eq. (\ref{newmar}) provides the following solutions

\begin{equation}
\phi_{vac}=\frac{1}{2}\left(1-\frac{\sigma^2}{2r}\right)\left(1\pm1\right).     
\end{equation}
The result for this previous equation gives a trivial non-degenerate vacuum $\phi_{vac}^{n=1}=0$, but it can also give the following non-trivial result 

\begin{equation}   \label{vacext}
\phi_{vac}^{n=1}=1-\frac{\sigma^2}{2r}.    
\end{equation}
Note that this result is the same obtained in \cite{OurEPL} when the BS eq. was analyzed. This is the case because in \cite{OurEPL}, the kinetic terms were ignored (Strong-field regime), which is precisely what happens in eq. (\ref{allpotentials}) when $n=1$. However, note that the result (\ref{vacext}) appeared for $n=2$ instead of $n=1$ in \cite{OurEPL}. The reason for this is that in \cite{OurEPL}, the derivative of the potential was considered instead of the potential itself as it is the case here. In any case, this only demonstrates that the regimes analyzed in this paper are more general than the simple case studied in \cite{OurEPL}. All the other results, with $n\neq0, 1$ are non-trivial. Eq. (\ref{allpotentials}) is by itself a potential function $V(\phi)$ and we can demonstrate that any derivative of this potential will not give any information different to the one obtained from eq. (\ref{allpotentials}). Then for example, if we find find the extreme condition for eq. (\ref{allpotentials}) as $\partial V(\phi)/\partial\phi=0$, we get

\begin{equation}   \label{Generalcondition}
\phi^2+\left(\frac{1}{2}\frac{\sigma^2}{r}-1\right)(n-1)\phi-\frac{\sigma^2}{2r}(n-1)(n-2)=0,
\end{equation}
which is just essentially the same result in eq. (\ref{allpotentials}) but with the coefficient terms shifted. Then for this case, when $n=1$, the previous equation gives a trivial result $\phi=0$. In this way, since any higher-order derivative provides the same information as the one obtained from eq. (\ref{allpotentials}), for the purpose of analysis, we will just focus on eq. (\ref{allpotentials}) for understanding the different regimes. Finally, note that although $\hat{H}_{BS}\phi^n=0$, still $\hat{p}\phi^n=n\phi^{n-1}\neq0$, considering $n\neq0$. Then the vacuum is only annihilated by the momentum if the field $\phi$ is itself trivial. Something which happens for some combination of parameters. Then for example, eq. (\ref{vacext}) gives a trivial result when $\sigma^2=2r$, which is the same condition for the Hamiltonian to be Hermitian. Then, at least for this particular case, there is a direct connection between flow of information and spontaneous symmetry breaking. When the martingale state is unique, no random fluctuations can change the prices in the market. However, when the martingale (vacuum) state is degenerate, then a multiplicity of possible ground states emerge and then any fluctuation generates changes in the information of the system and as a consequence, changes on the prices of the Option under analysis.  

\subsubsection{Vacuum conditions for $n\neq0, 1$: Weak and strong field regimes}

{\bf 1). Weak-field regime: $\phi^n<<\phi$}\\\\

For this regime, from eq. (\ref{allpotentials}), we obtain 

\begin{equation}
-\frac{\sigma^2}{2}(n-1)+\left(\frac{1}{2}\sigma^2-r\right)\phi\approx 0, 
\end{equation}
which can be easily solved to obtain

\begin{equation}   \label{Poke}
\phi_{vac}\approx\frac{\sigma^2}{\sigma^2-2r}(n-1).    
\end{equation}
Note that this result is trivial for $n=1$ for any combination of parameters. This means that in these situations, again random fluctuations cannot change the prices in the market. 
The case (\ref{Poke}) is also trivial when $\sigma^2=0$ (zero volatility). Zero volatility means zero fluctuations and then zero possibility of changing the prices in the market.   

{\bf 2). Strong Field regime:$\phi^n>>\phi$}\\\\

In this regime, eq. (\ref{allpotentials}) gives the following condition

\begin{equation}
r\phi^2+\left(\frac{1}{2}\sigma^2-r\right)n\phi\approx0.    
\end{equation}
We have two solutions; one of them is the trivial $\phi_{vac}\approx 0$ and the second solution is

\begin{equation}
\phi_{vac}=\left(1-\frac{1}{2}\frac{\sigma^2}{r}\right)n.
\end{equation}
Here again, this result is trivial if $\sigma^2=2r$, which is again the condition for the Hamiltonian to be Hermitian and then preserve information. Then it seems that this is the combination of parameters which avoids changes of information due to random fluctuations. Before going to the next section, we have to remark that the results obtained for the BS case will be valid for the MG case because the standard definition of martingale is independent of the stochastic volatility. The only change to be done for the solutions inside the MG equation, when we consider $\phi_{vac}$, is in the definition of the volatility $\sigma^2=e^y$, with $y$ representing the variable connected to the stochastic volatility.   

\section{A more general condition for the symmetry breaking in the Merton Garman equation}   \label{Secnew}

When we analyze the MG equation, the martingale condition is normally taken such that it is independent on the stochastic volatility. The volatility is a function of the variable $y$, as it is defined in eq. (\ref{range}). If the martingale condition is taken as independent of $y$, then any term with derivative with respect to this variable, will annihilate the state $S(x, t)$ defined previously. From this perspective, $y$ can be taken as fixed when we are determining the vacuum conditions. In this portion of the paper, we would like to define a more general martingale condition, such that the possible changes in $y$ can be considered. We will take the martingale state as

\begin{equation}   \label{newmartin}
\hat{H}_{MG}e^{x+y}=\hat{H}_{MG}S(x, y, t)=0.    
\end{equation}
Here we extend the arguments showed in eq. (\ref{range}), considering the extensions of the original martingale state $S(x, t)=e^x$. The condition (\ref{newmartin}) will be considered here as the extended martingale condition with $S(x, y, t)=e^{x+y}$. By using the Hamiltonian (\ref{MGHamilton}) and the result (\ref{newmartin}), we can observe that the condition for the Hamiltonian to annihilate the martingale state is

\begin{equation}   \label{corona4}
\lambda+e^y\left(\mu+\frac{\zeta^2}{2}e^{2y(\alpha-1)}+\rho\zeta e^{y(\alpha-1/2)}\right)=0, 
\end{equation}
as far as $e^{x+y}\neq0$, avoiding then the trivial solution. This previous condition is necessary for the state $e^{x+y}$ to be considered as the martingale state and it represents a constraint among the free-parameters involved in the MG equation.  

\subsection{The extended martingale condition and the flow of information}

Previously, when we studied the ordinary martingale condition, we could demonstrate that it can be also considered as a vacuum state. The vacuum in the BS case came out to be single if there is no flow of information through the boundaries of the system and it was degenerate if the information flows through the boundaries of the system. When the vacuum is single, the momentum, defined as the generator of the changes in prices, is a perfect symmetry. On the other hand, when the vacuum is degenerate, then the same symmetry is spontaneously broken because although the Hamiltonian annihilates the ground state (martingale condition), the momentum does not do it. The interesting point here is the connection between spontaneous symmetry breaking and flow of information, which is connected with the changes on the prices in the market \cite{3}. Something which has been suggested before in \cite{Winchi} in a different context. By using the previously proposed extended Martingale state, we can express it as a function of Quantum fields in the form $S(x, y, t)=e^{x+y}=\left(\sum_{n=0}^\infty\phi_x^n\right)\left(\sum_{m=0}^\infty\phi_y^m\right)$. By replacing this expression in eq. (\ref{MGHamilton}), we get

\begin{eqnarray}  \label{MGHamiltonMartingale}
-\frac{e^y}{2}n(n-1)\phi_x^{n-2}\phi_y^m-\left(r-\frac{e^y}{2}\right)n\phi_x^{n-1}\phi_y^m-\left(\lambda e^{-y}+\mu-\frac{\zeta^2}{2}e^{2y(\alpha-1)}\right)m\phi_x^n\phi_y^{m-1}\nonumber\\
-\rho\zeta e^{y(\alpha-1/2)}nm\phi_x^{n-1}\phi_y^{m-1}-\zeta^2 e^{2y(\alpha-1)}m(m-1)\phi_x^n\phi_y^{m-2}+r\phi_x^n\phi_y^m=0.
\end{eqnarray}
The vanishing condition comes out from the definition of martingale. The previous result is valid if $d\phi_x/dx=\sum_n\phi_x/n$ and $d\phi_y/dy=\sum_n\phi_y/n$. The equation (\ref{MGHamiltonMartingale}) is a quadratic equation for both fields, namely, $\phi_x$ and $\phi_y$. Then we can decide to solve it with respect to either field if convenient. In this paper we will not do it; we will rather analyze a few interesting cases and regimes in order to visualize the structure of the extended martingale condition for the MG equation. With the new definition of martingale, it is known that the symmetries under changes of price as well as the symmetries under changes of volatility are both spontaneously broken. This means that the information of the system changes due to fluctuations in the prices as well as fluctuations on the volatility. These situations are represented Mathematically as \cite{OurEPL}

\begin{equation}
\hat{p}_x\vert S>\neq0,\;\;\;\;\;\hat{p}_y\vert S>\neq0.    
\end{equation}
with $\vert S>$ representing the ket related to the martingale state $S(x, y, t)$. Additionally, $<x, y\vert\hat{p}_x\vert S>=\partial_x S(x, y, t)$ and $<x, y\vert\hat{p}_y\vert S>=\partial_y S(x, y, t)$; representing in this way the action of the operators $\hat{p}_{x, y}$ as the generators of translations in the prices and volatility related to the options. We can now analyze some important cases related to the condition (\ref{MGHamiltonMartingale}). Note for example that for $n=m=0$, we do not get anything interesting from this equation. Indeed, for this case, $r=0$ (zero interest rate). On the other hand, we can analyze the following interesting cases

\subsubsection{Extended martingale with n=0 and m=1}

For this case, eq. (\ref{MGHamiltonMartingale}) gives us the following result

\begin{equation}
\phi_{y\;vac}=\frac{1}{r}\left(\lambda e^{-y}+\mu-\frac{\zeta^2}{2}e^{2y(\alpha-1)}\right),    
\end{equation}
with an arbitrary value for $\phi_{x\;vac}$. This means that for this particular case, the symmetry under changes in the volatility is spontaneously broken, except when the condition 

\begin{equation}   \label{informationflowmama}
\lambda e^{-y}+\mu-\frac{\zeta^2}{2}e^{2y(\alpha-1)}=0,    
\end{equation}
is satisfied. This condition complemented with eq. (\ref{corona4}), gives the result

\begin{equation}
e^{y(\alpha-3/2)}=-\frac{\rho}{\zeta},    
\end{equation}
which is a constraint related to important parameters connected to the stochastic volatility \cite{3}. Note that in this case, the field $\phi_x$ is arbitrary. This means that the symmetry under changes of the prices is also spontaneously broken. Note in addition that the condition (\ref{informationflowmama}) corresponds to the condition for recovering the Hermiticity with respect to the variable $y$. As a consequence of this, the unique vacuum condition with respect to $y$, here represented by $\phi_{y\;vac}=0$, is connected to the no-flow of information with respect to volatility variable $y$. This means that when (\ref{informationflowmama}) is valid, then the random fluctuations coming from the stochastic volatility do not affect do not produce changes on the information of the system. Still for this case however, the information is not completely preserved due to the random fluctuations on the prices, which disappear when $r=e^y/2$, eliminating in this way all the non-Hermiticities in the MG Hamiltonian in eq. (\ref{MGHamilton}).     

\subsubsection{Extended martingale with n=1 and m=0}

For this case, eq. (\ref{MGHamiltonMartingale}) gives us the solution

\begin{equation}   \label{stvac}
\phi_{x\;vac}=1-\frac{e^y}{2r},    
\end{equation}
with $\phi_y$ arbitrary. Compare eq. (\ref{stvac}) with the case $n=1$ of the BS result in eq. (\ref{vacext}). The result is basically the same but with the re-defined volatility as a stochastic variable $\sigma^2=e^y$. Here again the single vacuum condition $\phi_{x\;vac}=0$ is related to the no-flow of information but this time with respect to the variable $x$ which is a function of the price of the stock. For this case, $\phi_{y\;vac}$ is arbitrary and then still the symmetries with respect to the changes in volatility are spontaneously broken. In this way, only the random fluctuations of the volatility affect the information content of the market. 

\subsubsection{Extended martingale with n=1 and m=1}

For this case, eq. (\ref{MGHamiltonMartingale}) gives us the solution

\begin{equation}   \label{Simplified Mart}
-\left(r-\frac{e^y}{2}\right)\phi_y-\left(\lambda e^{-y}+\mu-\frac{\zeta^2}{2}e^{2y(\alpha-1)}\right)\phi_x
-\rho\zeta e^{y(\alpha-1/2)}+r\phi_x\phi_y=0.    
\end{equation}
We can solve this expression for either Quantum field, $\phi_y$ or $\phi_x$. One interesting situation emerges when the Hermiticity condition for the Hamiltonian (\ref{MGHamilton}) is satisfied. This occurs when eq. (\ref{informationflowmama}) is satisfied and when $e^y=2r$. In such a case we have from eq. (\ref{Simplified Mart})

\begin{equation}   \label{rhobebe}
\phi_{x\;vac}\phi_{y\;vac}=\frac{\rho\zeta e^{y(\alpha-1/2)}}{r}.
\end{equation}
If in addition $\rho=0$ or $\zeta=0$, then we would have $\phi_x\phi_y=0$, and then either, the symmetry is spontaneously broken with respect to translation in prices but not with respect the variations in volatility or on the contrary, the symmetry is spontaneously broken with respect to changes in the volatility but we still have a perfect symmetry with respect to changes in the prices of the options. The same result also suggests that the simultaneous changes in the price with a subsequent changes in volatility (and viceversa), represent a perfect symmetry for the system as far as $\rho=0$ or $\zeta=0$. This means that for this combined symmetry, simultaneous random fluctuations of the volatility and prices keep the information of the system unchanged. Note that $\rho$ is a variable representing the correlation between the Gaussian noises $R_1$ and $R_2$ as it was defined in eq. (\ref{corona3}). Then $\rho=0$ means zero correlation between the Gaussian noises. On the other hand, if $\zeta=0$, then the Gaussian noise $R_2$ decouples and it comes out that it does not affect the changes in the stochastic volatility. This last issue can be observed from eq. (\ref{corona5}).       

\subsubsection{Strong Field condition $\phi_x^n>>\phi_x$ and $\phi_y^m>>\phi_y$; for any value of $n$, $m$}

For this case, from eq. (\ref{MGHamiltonMartingale}), we obtain

\begin{equation}   \label{meetingagain}
\phi_{x\;vac}\phi_{y\;vac}\approx \left(1-\frac{e^y}{2r}\right)n\phi_{y\;vac}+\left(\frac{\lambda}{r} e^{-y}+\frac{\mu}{r}-\frac{\zeta^2}{2r}e^{2y(\alpha-1)}\right)m\phi_{x\;vac},    
\end{equation}
which can be approximated to zero under the weak-field approximation if we ignore cubic order terms in eq. (\ref{MGHamiltonMartingale}) in comparison to the fourth-order terms. Even if we do not ignore such terms, eq. (\ref{meetingagain}) is exactly zero under the Hermiticity conditions (no-flow of information) defined by eq. (\ref{informationflowmama}) and the constraint $e^y=2r$. In any case, this result shows that for strong field approximation for both fields, connected to the symmetries related to changes in prices and volatility, it is not possible to have a single vacuum condition related to both symmetries. However, it is possible to have a perfect vacuum condition after the simultaneous application of both symmetries, namely, the symmetries under changes in the price plus the symmetries under changes in volatility. This result comes out to be similar to the one analyzed in eq. (\ref{rhobebe}), if we have $\rho=0$ or $\zeta=0$ as it was explained previously.  

\subsubsection{Weak-field condition $\phi_x^n<<\phi_x$ and $\phi_y^m<<\phi_y$; for any value of $n$, $m$}

For this regime, eq. (\ref{MGHamiltonMartingale}) gives the approximation

\begin{equation}
-\frac{e^y}{2}n(n-1)\phi_y^2-\rho\zeta e^{y(\alpha-1/2)}nm\phi_x\phi_y-\zeta^2 e^{2y(\alpha-1)}m(m-1)\phi_x^2\approx0,    
\end{equation}
which for $n=m=1$ would give again $\phi_{x\;vac}\phi_{y\;vac}\approx0$ as before. We can certainly solve this previous equation as a quadratic equation in either, $\phi_x$ or $\phi_y$. If we decide to solve for $\phi_y$, we get

\begin{equation}
\phi_{y\;vac}\approx \frac{\rho\zeta e^{y(\alpha-3/2)}m\phi_{x\;vac}}{(1-n)}\left(1\pm\sqrt{1-\frac{2(m-1)(n-1)}{\rho^2nm}}\right).    
\end{equation}
Note the coupling between $\phi_{y\;vac}$ and $\phi_{x\;vac}$, such that the triviality of one field implies the triviality of the other. However, it is still possible to have a trivial value for $\phi_{y\;vac}$ and still a non-trivial one for $\phi_{x\;vac}$ if $\zeta=0$ or if $m=1$. The coupling between fields means that the broken symmetry with respect to the changes in the prices is connected to the broken symmetry with respect to changes in volatility. This also means that the random fluctuations of the prices are tied to the random fluctuations of the volatility. This naturally is connected to the flow of information inside the system.

\subsubsection{Weak-Strong-field approximation: $\phi_x^n>>\phi_x$ and $\phi_y^m<<\phi_y$}

For this case, eq. (\ref{MGHamiltonMartingale}) becomes

\begin{equation}
\left(\lambda e^{-y}+\mu-\frac{\zeta^2}{2}e^{2y(\alpha-1)}\right)\phi_x^2\phi_y-\zeta^2e^{2y(\alpha-1)}(m-1)\phi_x^n\phi_y^{m-2}\approx0,    
\end{equation}
which gives a trivial solution for $\phi_{x\;vac}\approx0$ and the following result for $\phi_{y\;vac}$

\begin{equation}
\phi_{y\;vac}\approx\frac{\zeta^2e^{2y(\alpha-1)}(1-m)}{\lambda e^{-y}+\mu-\frac{\zeta^2}{2}e^{2y(\alpha-1)}}.    
\end{equation}
Then for this case, the symmetry under changes in the prices is not broken but the one related to the changes in volatility is spontaneously broken, except when $m=1$ or $\zeta=0$. Then in the most general situations, the random fluctuations cannot affect the prices of the market under this approximation. Still, the random fluctuations generate changes in the information content of the system because the location of the equilibrium condition changes due to the random fluctuations on the volatility. Once again, a value of $\zeta=0$ decouples the Gaussian noise from the evolution in time of the volatility.  

\subsubsection{Weak-Strong-field approximation: $\phi_x^n<<\phi_x$ and $\phi_y^m>>\phi_y$}

For this case, eq. (\ref{MGHamiltonMartingale}) is approximated to 

\begin{equation}
\frac{e^y}{2}(n-1)\phi_y^2+\left(r-\frac{e^y}{2}\right)\phi_x\phi_y^2\approx0.    
\end{equation}
This equation gives a trivial result for $\phi_{y\;vac}\approx0$ and a non-trivial one for $\phi_{x\; vac}$ as

\begin{equation}
\phi_{x\;vac}\approx\frac{(1-n)\frac{e^y}{2}}{r-\frac{e^y}{2}},
\end{equation}
which vanishes when $n=1$ or when $y\to-\infty$ ($\sigma^2\to0$). Then for this case, in general, the symmetry under changes of the volatility is a perfect symmetry, meanwhile, the symmetry under changes of the prices is spontaneously broken. Then the random fluctuations only affect the the information contained on the prices in the market in these special circumstances. Interestingly, when $y\to\infty$, the previous expression becomes

\begin{equation}
\phi_{x\;vac}\approx n-1,
\end{equation}
which is independent from any parameter and only depends on the order of the term in the series expansion under analysis. 

\section{Conclusions}   \label{Sec5}

In this paper we have analyzed the degeneracy of the martingales state on regimes which were not considered before in previous papers. We have also connected the concept of spontaneous symmetry breaking in Quantum Finance with the flow of information through the market. The analysis has been done for both, the BS equation as well as for the case where the volatility is stochastic, namely, for the MG equation.    
For the BS equation as well as for the MG one, the symmetries under changes of the prices of the Options cannot annihilate the vacuum (taken as the martingale state). These symmetries are then spontaneously broken in both cases. This means that random fluctuations in the market generate changes in the information of the system, forcing it to move toward another equilibrium condition. We found the conditions where the vacuum, in these two situations (MG and BS equations), becomes unique. These come out to be the same conditions for which there is no flow of information through the boundaries of the system (Market), or equivalently, the same conditions  correspond to the cases where the Hamiltonian becomes Hermitian. Hermitian Hamiltonians in general preserve the information of the system. Under the ordinary definition of martingale, the connection between flow of information and spontaneous symmetry breaking, is exact. We could then extend the martingale condition for including the symmetries under changes of the stochastic volatility in a similar fashion to what is done in \cite{OurEPL}. We defined then the momentum for the volatility as the generator of these transformations. We could also find the conditions under which the martingale state (vacuum) becomes unique for different regimes and for the different series expansion terms. A unique vacuum or martingale state, means that the random fluctuations cannot change the information of the system under such special situations. In \cite{OurEPL} the authors only focused on the strong field approximation. Here however, we could complete the scenario by analyzing the weak-field approximation for the BS as well as for the MG case. In the weak-field approximation, the kinetic terms are the most relevant. Since the kinetic terms, expressed as a function of a Quantum Field, behave as potential terms; then their existence affects the vacuum location and behavior. For the case of the MG equation, we have to define two different Quantum Fields, one corresponding to the prices in the market and the other related to the volatility. In this paper we analyzed the regimes where both fields in the MG equation are either, strong or weak. Finally, we analyzed the situations where one field is strong and the other is weak. In these cases, it came out that one of the symmetries involved were perfect but the other one was spontaneously broken in general, except for some specific combination of parameters. Perfect symmetries mean conservation of the information under random fluctuations. This is a very important statement for the stock market analysis.  \\\\

{\bf Acknowledgement}
I. A. would like to thank to Prof. Yu Chen from the University of Tokyo for providing useful references and bibliography, as well as very useful discussions.

\bibliographystyle{unsrt}  
%\bibliography{references}  %%% Remove comment to use the external .bib file (using bibtex).
%%% and comment out the ``thebibliography'' section.

%%% Comment out this section when you \bibliography{references} is enabled.

\end{document}